# The Force Field for Amino Acid Based Ionic Liquids: Polar Residues

Eudes Eterno Fileti and Vitaly V. Chaban

[1] Instituto de Ciência e Tecnologia, Universidade Federal de São Paulo, 12231-280, São José dos Campos, SP, Brazil

**Abstract**. Ionic liquids (ILs) constitute one of the most active fields of research nowadays. Many organic and inorganic molecules can be converted into ions via relatively simple procedures. These ions can be combined into ILs. Amino acid based ILs (AAILs) represent a specific interest due to solubilization of biological species, participation in enzymatic catalysis, and capturing toxic gases. We develop a new force field (FF) for the seven selected AAILs comprising 1-ethyl-3-methylimidazolium cation and amino acid anions with polar residues. The anions were obtained via deprotonation of carboxyl group. We account for peculiar interactions between the anion and the cation by fitting electrostatic potential for an ion pair, in contrast to isolated ions. The van der Waals interactions were transferred from the CHARMM36 FF with minor modifications, as suggested by hybrid density functional theory. Compatibility between our parameters and CHARMM36 parameters is preserved. The developed interaction model fosters computational investigation of ionic liquids.

**Introduction**

Ionic liquids (ILs) play an important role in numerous chemically, physically, and biologically relevant technological processes.[1-8] Due to versatile chemical nature and low melting points, ILs constitute excellent solvents for a variety of exciting applications.[5, 9-20] The physical chemical properties of ILs are important for capturing toxic gases, separation applications through biphasic systems, solubilization of nanoscale entities, such as nanotubes and fullerenes.[1, 12, 13, 21-23]

Amino acid (AA) ionic liquids (AAILs) are a relatively new research field.[24] They can be obtained from amino acids and bulky cations, such as imidazolium-based and phosphonium-based ones.[24, 25] Alternative cations can be also employed, as long as the resulting entity is environmentally stable and melts at significantly low temperature. Importantly, many of the proposed AAILs exhibit definitely large liquid state temperature ranges distinguishing them from conventional molecular solvents. These temperature ranges can be further tuned by addition of other – molecular or ionic – co-solvents.[20, 24, 26, 27]

AAILs are currently being probed for enzymatic reactions, adsorption of carbon and sulfur dioxides, solubilization of proteins and other biologically relevant macromolecules.[5, 8, 28] The comprehensive force fields (FFs) for these ionic liquids are not available yet. However, computational investigation is a powerful and often reliable tool offering atomistic-precision physical insights.[29-31] Reliability and internal compatibility of the force field are a central issue in view of efficient computer simulations. Another important issue is constituted by a computational cost of the FF. Although a wide variety of complicated models can be, in principle, derived, most of them are impractical due to (1) huge computational cost; (2) imperfect parallelization (scalability). The more mathematically sophisticated interaction model is supplied, the harder implementation of a reasonable parallelization appears. In turn, real-life applications require simulation of the systems containing at least thousands of interaction centers, while characteristic time scales of most interesting processes range from nanoseconds to microseconds. Observation of phase transitions, i.e. crystallization, requires even larger simulated times, up to seconds. Simulation of metastable systems, e.g. supersaturated solutions, superheated liquids, etc, necessitates comparable sampling. Simple, but realistic atomistic-precision and coarse-grained models are important to handle such physical problems.

We develop an additive force field (Coulomb attraction / repulsion plus weak van der Waals attraction to describe non-bonded interactions) for AAILs with polar residues. In the case of lacking experimental data, we rely on our previous experience in deriving FFs for ILs and electronic structure calculations. We account for specific cation-anion interactions (partial charge transfer) by tuning point electrostatic charges and the location of Lennard-Jones potential depth. Although the proposed FF is formally additive (non-polarizable), it captures all major atomistic-level effects, which are responsible for quantitatively correct thermodynamics, structure, and transport properties. The performance of the models for all AAILs featuring polar residues was tested using classical atomistic-precision molecular dynamics (MD) simulations.

**Methodology**

Electronic structure description of the AAIL ion pairs was carried out using omega B97XD hybrid density functional theory (DFT) functional.[32] This functional exhibits an improved performance for electronic energy levels, non-covalent interactions (an important issue) and thermochemistry. Therefore, B97XD was preferred over more traditional DFT functionals. The wave function was expanded using a set of gaussian basis functions provided in the 6-311G basis set. Additionally, the polarization and diffuse functions have been centered on each heavy atom (6-311+G*) to provide an improved description of the ions containing high-energy valence electrons (as opposed to neutral molecules). Polarization functions are important to correctly describe an electronic density of anions. The number of basis functions per system varied with respect to the total number of electrons. No pseudopotentials were used for the core electrons. The selected basis set is considered to provide a trustworthy approximation of the real wave function.

The electrostatic potential (ESP) was computed using the Möller-Plesset second-order perturbation theory, MP2, with precisely the same set of basis functions as in omega B97XD. The reason of preferring MP2 over DFT is a better compatibility with the CHARMM[33, 34] family of force fields. The ESP was approximated using a set of point charges. For simplicity, each charge was centered at an atom, including all hydrogen atoms. The ChelpG scheme with a default grid size in Gaussian 09 (*www.gaussian.com*) was employed to perform a charge assignment.[35, 36] All nuclear geometries (isolated ions and ion pairs) were optimized to correspond to the local minimum configuration of the electron-nuclear system. The electronic structure computations were performed in Gaussian 09, revision D (*www.gaussian.com*).[36]

Certain AAILs from the selected set exhibit very high viscosities at room conditions. This is due to the large and polar AA residues. We enhance sampling by performing all MD simulations at 500 K. Heat of vaporization is given by the difference between potential energy in condensed phase and potential energy of an isolated cation-anion pair, normalized per mole of ion pairs. Note that the *RT* product must be added to the result of the energy subtraction, whereas *T* must be set to 500 K. Mass density is recomputed from the variation of unit cell volume over equilibrium MD simulation, since the mass of the simulated cell remains constant. Diffusion constant is provided by the slope of mean-square displacements vs. simulated time. Shear viscosity is estimated by the non-equilibrium MD simulations from energy dissipation upon cosine-shape accelerations. The method is in details described by Hess elsewhere.[37] The list of the simulated systems is provided in Table 1.

Table 1. The AAIL systems considered in the present work. The quantity of ion pairs per AAIL was selected with respect to the cation and anion sizes (in number of atoms). We perform a very large sampling for the shear viscosity calculations to minimize standard deviations of the resulting values

| AAIL | Number of ion pairs | Number of interaction centers | Equilibrium MD (ns) | Non-equilibrium MD (ns) |
|---|---|---|---|---|
| [emim][asn] | 150 | 5250 | 20 | 100 |
| [emim][cys] | 175 | 5600 | 25 | 110 |
| [emim][gln] | 150 | 5700 | 25 | 110 |
| [emim][gly] | 200 | 5600 | 25 | 110 |
| [emim][pro] | 150 | 5250 | 20 | 100 |
| [emim][ser] | 175 | 5600 | 25 | 110 |
| [emim][thr] | 150 | 5250 | 20 | 100 |

The Cartesian coordinates were saved every 5 ps. The thermodynamic quantities and their components, where applicable, were saved every 0.02 ps. More frequent saving of trajectory components was preliminarily tested, but no systematic accuracy improvement was found. The equilibration of all MD systems was performed during 10.0 ns (equilibrium MD simulations). The basic thermodynamics quantities (unit cell volume, energy terms) were used to record the achieved equilibrium. Similarly, the constant flow in the case of non-equilibrium MD was established during 10.0 ns.

All simulations were carried out in the constant-pressure constant-temperature ensemble. The equations of motion were propagated with a time-step of 2.0 fs. Such a relatively large time-step was possible due to constraints imposed on the carbon-hydrogen covalent bonds (instead of a harmonic potential). The electrostatic interactions were simulated using direct Coulomb law up to 1.3 nm of separation between the interaction sites. The electrostatic interactions beyond 1.3 nm were accounted for by computationally efficient Particle-Mesh-Ewald (PME) method.[38] It is important to use PME method in case of ionic systems, since electrostatic energy beyond the cut-off usually contributes 40-60% of total electrostatic energy. The Lennard-Jones-12-6 interactions were smoothly brought down to zero from 1.1 to 1.2 nm using the classical shifted force technique. The constant temperature (500 K) was maintained by the Bussi-Donadio-Parrinello velocity rescaling thermostat[39] (with a time constant of 0.5 ps), which provides a correct velocity distribution for a statistical mechanical ensemble. The constant pressure was maintained by Parrinello-Rahman barostat with a time constant of 4.0 ps and a compressibility constant of $4.5\times10^{-5}$ bar$^{-1}$. All molecular dynamics trajectories were propagated using the GROMACS simulation suite.[40] Analysis of thermodynamics, structure, and transport properties was done using the supplementary utilities distributed with the GROMACS simulation suite,[40] where possible, and the in-home tools.

Each AAIL was placed in the cubic periodic MD box (Figure 1), whose density was calculated to approximately correspond to ambient pressure at the requested temperature (500K).

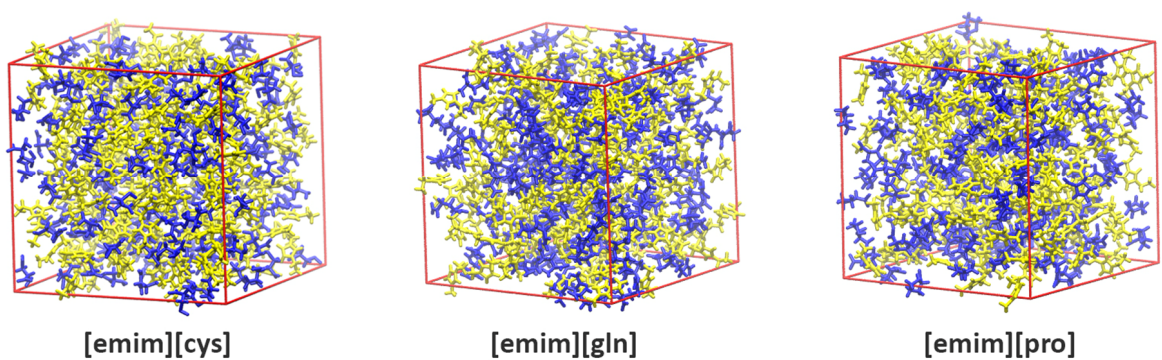

[emim][cys]  [emim][gln]  [emim][pro]

Figure 1. Molecular dynamics unit cells for selected AAILs. The imidazolium cations are in yellow and the amino acid anions are in blue. The abbreviations of anions in this figure and throughout our discussion coincide with the abbreviations of the corresponding AAs.

**Force Field Derivation**

Figure 2 depicts ion pair configurations corresponding to minimized internal energy of the system. The most positively charged hydrogen atom is coordinated by carboxyl group of all cations. Carboxyl oxygen carries a strong excess electron charge (ESP charge of -0.80e). Its pronounced affinity to an imidazole hydrogen atom can be described as an energetically driven desire to get protonated. Hydrogen bonding is mainly of electrostatic nature. Additionally, the optimized structure of these ion pairs is confirmed by classical MD simulations employing the CHARMM36 force field.

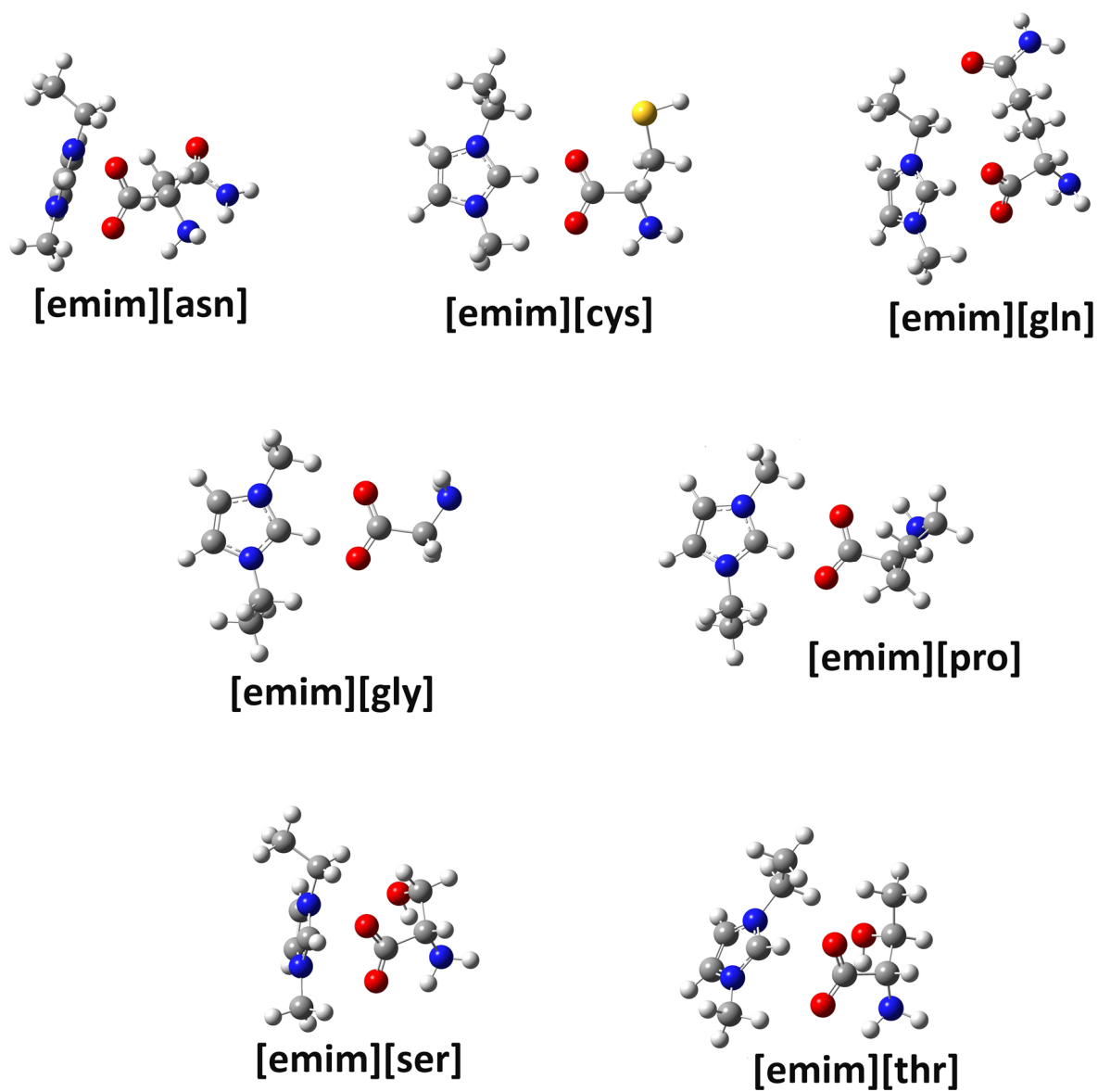

Figure 2. Hybrid DFT optimized structures for each of the ion pair investigated. Note, the anions are abbreviated using standard notation for the corresponding amino acids. The anions were obtained by deprotonation of carboxyl groups of these AAs.

It is an important feature of AAILs with polar residues, discovered in this work, that no residue participates in binding with the imidazolium cation. Well-defined cation-anion coordination implies prevalence of contact ion pairs in all seven AAILs. If these AAILs are dissolved in water or other polar solvent, the cation-anion pairs must dominate the solution structure. In turn, solubility in nonpolar solvents and solvents of weak polarity is prohibited by the strong hydrogen bonding. Generally, strong binding of cation or anion is not favorable for ionic liquids, since it increases melting temperature. Additionally, strong binding increases viscosity. Industrial applications of this sort of AAILs will be most successful in combination with the low-viscous polar solvents.

According to hybrid DFT, electronic polarization takes place between the COO⁻ moiety of the anion and imidazole ring of the cation. This effect brings non-additivity into the non-bonded interaction potential. Therefore, the CHARMM36 parameters cannot describe these systems perfectly. The difference between various amino acid anions, in terms of assigned ESP point charges (Table 2), does not exceed the expected difference for various anionic conformations during molecular dynamics at finite temperature. Due to this feature, the refinement procedure can be unified to account only for carboxyl group of the anion. Amino group is not sensitive to the presence of the imidazole ring.

The simplest amino acid can be used to derive charges. We selected 1-ethyl-3-methylimidazolium alanine, [emim][ala], ion pair to derive point electrostatic charges for the corresponding moieties of all seven AAILs. The ion pair geometry was re-optimized using MP2 method and ESP potential was fitted on the basis of the highly accurate MP2 electron density. The obtained charges amount to -1.17e (nitrogen of amino group), +0.38e (hydrogen of amino group), -0.80e (oxygen of carboxyl group), +0.75e (α-carbon).

The ESP charge located on the imidazole moiety in the ion pair configuration equals to +0.85e (in contrast to +1e in the isolated ion configuration). The charge deficiency is delocalized over the imidazole ring. It cannot be ascribed to any particular interaction center. For this reason, we uniformly scaled down imidazole point charges. The original charges on [emim]⁺ were obtained at the MP2/6-311+G* level of theory for the isolated cation to preserve isolated cation symmetry. These charges were multiplied by 0.85 and rounded to two decimal digits. To recapitulate, the charge re-assignment was only done for the imidazole ring of the cation and the coordination center of the anions. All other charges were taken from the CHARMM36 force field, as these models for AAs were well trained previously.

The CHARMM36 FF slightly overestimates the hydrogen bond length (Table 2) between carboxyl group oxygen atom and hydrogen atom of imidazole ring, as compared to the results of hybrid DFT functional, omega B97XD. We changed the three cross-sigma Lennard-Jones parameters for the hydrogen bonds (oxygen of anion – hydrogen of cation) in those AAIL, where the discrepancy between CHARMM36 and omega B97XD. These AAILs are [emim][cys], [emim][gly], and [emim][pro] (see Table 2). Note, the CHARMM36 FF predicts the same hydrogen bond length, 0.21 nm, in all AAILs. Ionic nature of AAILs favors hydrogen bonding. The discovered differences must be accounted for to obtain more precise simulated densities. Densities, in turn, are responsible for adequate reproduction of a number of other structure and transport properties of the liquid-matter systems.

Table 2. The length of the cation-anion hydrogen bonds obtained using the hybrid DFT functional, omega B97XD. Note, these results do not account for thermal motion

| [emim][asn] | [emim][cys] | [emim][gln] | [emim][gly] | [emim][pro] | [emim][ser] | [emim][thr] |
|---|---|---|---|---|---|---|
| H-bond length, nm | | | | | | |
| 0.209 | 0.172 | 0.181 | 0.173 | 0.175 | 0.201 | 0.186 |

**Results and Discussion**

This section discusses simulated thermodynamics (Table 3), structure (Figures 3-4), and transport properties (Table 3) of the seven simulated AAILs. The experimental data for this set of ionic liquids are not available yet. However, we applied a well-established phenomenological method for cation-anion specific binding. The good performance of this method was proven for a variety of room-temperature ionic liquids representing pyridinium and imidazolium families. The simulated AAILs can be compared to one another to derive qualitative insights into this novel class of ionic compounds.

The densities (Table 3) of all the simulated amino acid based ionic liquids range from 980 kg m$^{-3}$ ([emim][pro]) to 1113 kg m$^{-3}$ ([emim][ser]). Note, the observed density differences are quite significant indicating that the physical chemical properties of these compounds are also quite different. This is despite obvious similarities in the cation-anion binding, see above. In general, the polar AAILs exhibit lower densities than the nonpolar ones.[30] It can be rationalized well, since polar entities interact more strongly with one another and, therefore, are more tightly packed.

The enthalpy of vaporization (Table 3) ranges from 128 kJ mol$^{-1}$ ([emim][gly]) to 161 kJ mol$^{-1}$ ([emim][asn]). The enthalpy of vaporization is systematically lower than that of nonpolar amino acids ionic liquids. The latter fluctuates between 157 kJ mol$^{-1}$ and 216 kJ mol$^{-1}$.[30] No experimental values for this important thermodynamic property are available thus far. This situation may indicate insufficient in the physical communicate to AAILs, which are mainly of interest for biophysical and biochemical applications. At the same time, it may indicate experimental complexities to measure this property. Based on all results of our study, we anticipate that melting and boiling points of many of these AAILs are quite high due to the strong cation-anion binding. The reported heats of vaporization (Table 3) are the first available data. They provide a good reference and inspiration for future experimental investigations of the amino acid ionic liquids.

Shear viscosity and ionic self-diffusion coefficients are in concordance with one another. Larger viscosity means smaller diffusion and vice versa. The most mobile AAIL is [emim][gly] with a cationic diffusion of 78 $\mu m^2 s^{-1}$ and an anionic diffusion of 69 $\mu m^2 s^{-1}$. In turn, shear viscosity at the same temperature amounts to 2.0 cP. Importantly, the cation is faster than the anion in all AAILs. In the case of larger AA anions, the cation is faster by more than two times. For instance, $D(+) = 29$ $\mu m^2 s^{-1}$, while $D(-) = 12$ $\mu m^2 s^{-1}$ in [emim][trh]. Note that all the considered AAIL are much more viscous at room temperature. Therefore, they were simulated at significantly higher temperature to avoid a wide range of possible methodological problems, such as insufficient sampling times.

Table 3. Properties of amino acid based ionic liquids obtained using the newly developed force field at 500 K. Standard deviation was obtained by averaging out over 10 trajectory blocks of 10 ns each.

| AAIL | $\rho$ (kg m$^{-3}$) | $\Delta H_{vap}$ (kJ mol$^{-1}$) | D(+) ($\mu$m$^2$ s$^{-1}$) | D(−) ($\mu$m$^2$ s$^{-1}$) | $\eta$ (AAIL), cP |
|---|---|---|---|---|---|
| [emim][asn] | 1113 | 161 | 30±2 | 17±2 | 7.2±0.1 |
| [emim][cys] | 1081 | 137 | 48±4 | 35±2 | 3.2±0.1 |
| [emim][gln] | 1090 | 155 | 27±2 | 14±1 | 8.0±0.2 |
| [emim][gly] | 1028 | 128 | 78±5 | 69±4 | 2.0±0.0 |
| [emim][pro] | 982 | 134 | 42±3 | 27±2 | 3.7±0.1 |
| [emim][ser] | 1077 | 145 | 38±3 | 19±1 | 6.4±0.2 |
| [emim][trh] | 1054 | 148 | 29±3 | 12±1 | 9.3±0.2 |

Figure 3 depicts radial distribution functions (RDFs) computed between centers-of-mass of cations and anions in all AA based ionic liquids. All cation-anion RDFs exhibit a well-defined peak at 0.5-0.6 nm. The height of this peak varies from 1.7 units in [emim][gln] to 2.2 units in [emim][gly]. It is interesting to note that heights of the first peak are very similar to one another in all seven AAILs. The detected similarity constitutes an indirect proof that atomistic-level arrangement in the condensed phase of AAILs with polar radicals is determined solely by the carboxyl group and the acidic hydrogen atom. The polar radicals participate poorly in the cation-anion binding. The second peak of the cation-anion RDFs is weak and localized at 1.0-1.1 nm. Both the second minimum and the third maximum are absent.

Interestingly, the anion-anion center-of-mass RDF indicates clear structuring with two maxima. These peaks are not as high as the cation-anion peaks in the corresponding compounds, which must be expected. The carboxyl group, which is responsible for a strong cation-anion binding, plays a major role in the coordination of anions. Partial electrostatic charge localized on the oxygen atom of the carboxyl group equals to -0.80e. It is compensated by a nearly symmetrically high charge of the nitrogen atom in the amino group, +1.17e. See force field derivation section for details. These charges generate strong attraction among anions of all AAILs in the condensed phase. In turn, coordination of cations is notably weak exhibiting only one tiny peak, since the cation-cation electrostatic interactions are much weaker than those of the AA anions.

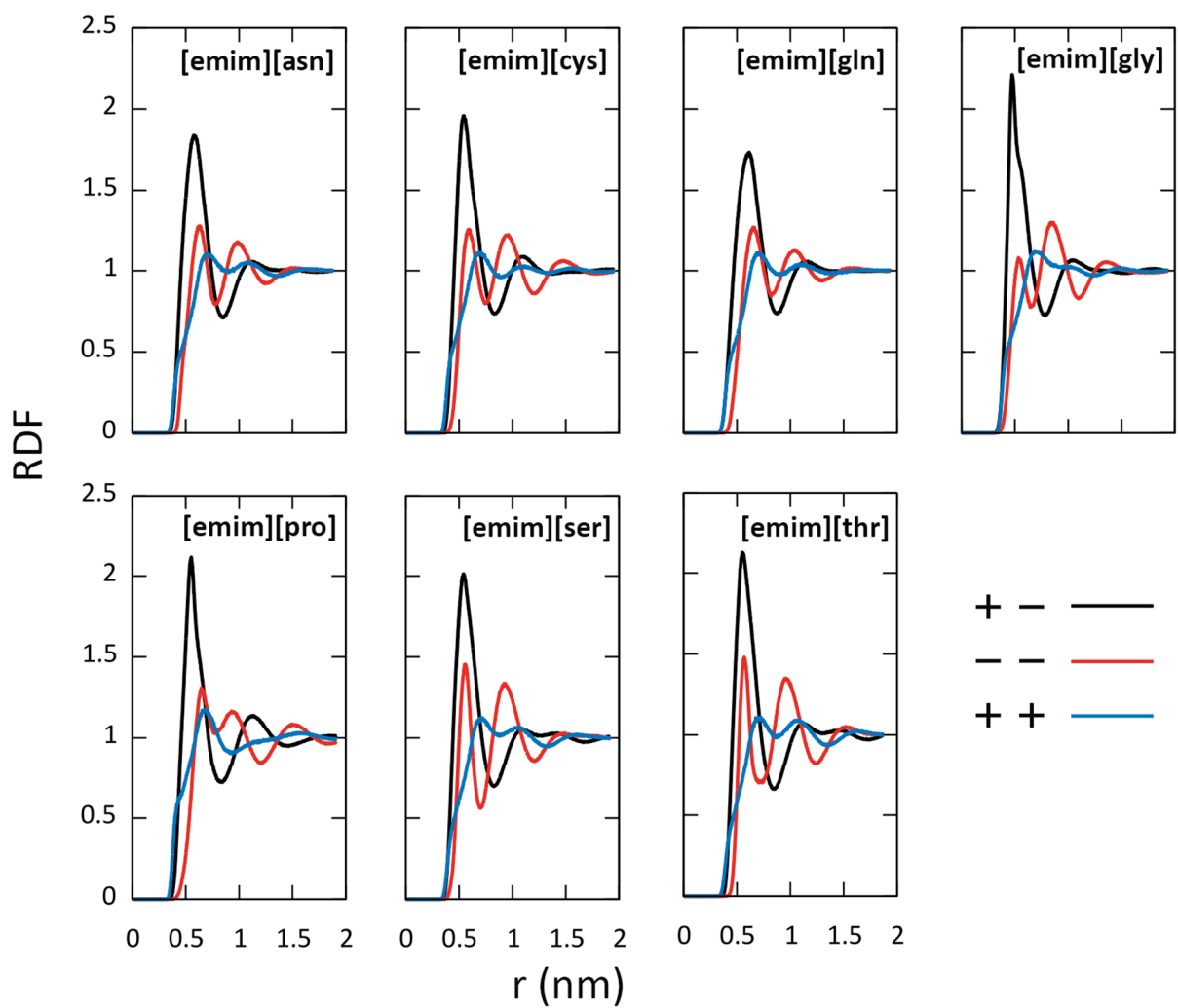

Figure 3. Ion–ion radial distribution functions computed between centers-of-mass of ions in AAILs. See legend for designations. These functions provide a probability of finding an ion pair at a certain distance compared to the average density of ions.

Figure 4 investigates hydrogen bonding atom pair, formed by an electron-rich oxygen atom of the carboxyl group and electron-deficient hydrogen atom of the imidazole ring. Compare the location of the first peak, 0.18-0.22 nm, with distances in Table 2. Note that distances in Table 2 do not include an effect of thermal motion at 310 K, while Figure 4 includes this effect. Accurate reproduction of the slight differences in the hydrogen bond lengths is possible thanks to the introduced refinements in the cross-sigma Lennard-Jones (12,6) interaction parameters. Unlike in the case of centers-of-mass RDFs (Figure 3), the heights of the peaks for different compounds are noticeably different. Larger lengths of the oxygen-hydrogen bonds nearly proportionally decrease the heights of the corresponding RDFs. Indeed, [emim][gln], [emim][ser], [emim][asn], and [emim][thr] exhibit approximately twice smaller first peaks than [emim][pro], [emim][cys], [emim][gly]. The second peak is located at ca. 0.6 nm in all AAILs, whereas the first minimum is not pronounced. Figure 4 indicates that hydrogen bonding plays a major role in maintaining the condensed-phase structure of the amino acid based ionic liquids. Hydrogen bonding appears in inverse proportion to ionic mobility. [emim][pro], [emim][cys], and [emim][gly] AAILs exhibit smaller shear viscosity and larger ionic self-diffusion coefficients. These same AAILs exhibit the highest peaks in the oxygen-hydrogen RDFs (Figure 4). These AAILs also possess somewhat spatially smaller anions than other considered AAILs. The size of the ion, along with its inter-ionic interactions, contributes to ion mobility as a whole, and therefore, energy dissipation (shear viscosity). The presence of the large AA residue weakens a hydrogen bond, since other moieties of the anion also participate in binding with cations. When these interaction trends are competitive, all the corresponding bonds become somewhat weaker. Weakening of hydrogen bonds upon molecular size increase is a known feature in chemistry. Indeed, the strongest hydrogen bonds are known in the condensed phases of small molecules, such as liquid hydrogen fluoride and water, while organic compounds form significantly weaker and longer hydrogen bonds.

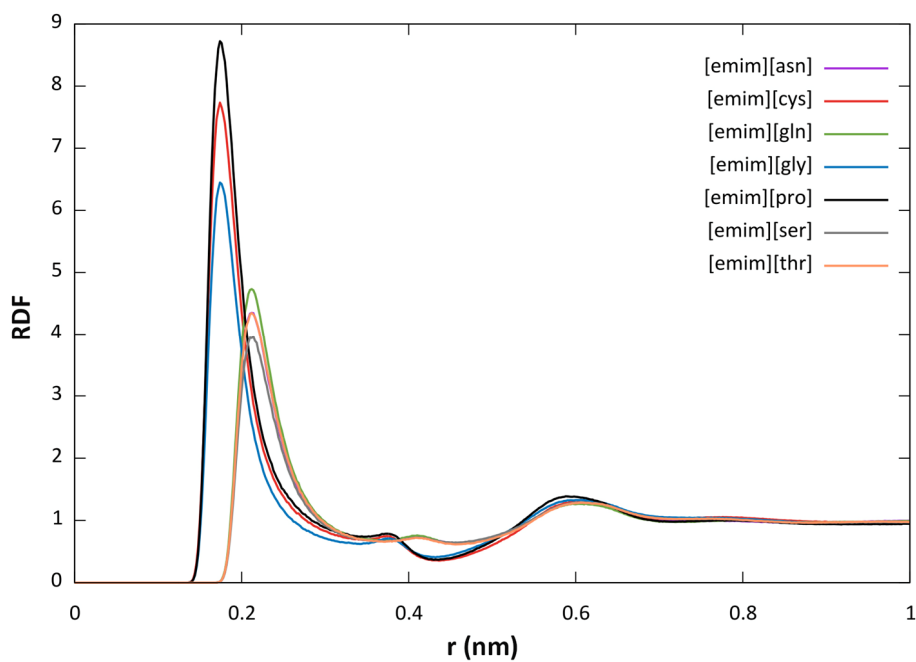

Figure 4. Radial distribution functions computed between the most positively charged atom of the imidazolium cation (ring hydrogen) and the most negatively charged atom of the anion (oxygen of COO⁻ amino acid moiety). Note that RDFs in [emim][asn] and [emim][thr] AAILs nearly coincide.

**Possible Pitfalls**

Theoretical predictions must be always thoroughly validated through experiment. In chemistry, experiment normally goes first, while theoretical explanation arrives later. There were bright exceptions to this rule in history though. Theoretical advances can foster and inspire experimental efforts. Alternatively, these two ways of understanding the Universe can develop in parallel without an explicit intersection. The experimentally determined ionic conductivities[24] of room temperature of AAILs are so small that the quantities of this order cannot be captured by MD simulations with a reasonable accuracy. We do not simulate them for this reason. Other experimental data for AAILs based on 1-ethyl-3-methylimidazolium cation are also not available thus far. However, this must not be an obstacle for the FF development, since the FF may itself foster the progress in the field. The present methodology was successfully employed in the case of the imidazolium-based and pyridinium-based ionic liquids,[29] where enough experimental data are available for comparison. It has been clearly shown that a reliable FF can be derived without an access to experimental data.

Why was a single ion pair used to reproduce electronic polarization in the condensed phase of AAILs? It is an obvious approximation, which should be taken with caution. Such an approximation was successfully used in Ref. [29], providing trustworthy predictions of structure and transport properties. Recall, only a single cation–anion coordination site is suggested by density functional theory calculations and subsequent molecular dynamics simulations. Therefore, namely an ion pair constitutes a major pattern in the condensed state structure of these AAILs.

Only a single cation–anion orientation is considered to compute electrostatic potential. ESP was subsequently fitted by a set of point charges. Using multiple heated configurations for charge derivation algorithm, one theoretically can improve an accuracy of the FF parameters. However, it is useful to remember that the accuracy of the computational model depends not only on the parameter set, but – first of all – on the functional forms of the model Hamiltonian. As long as a simple, computationally efficient, additive scheme is employed, the results cannot compete in accuracy with the electronic structure methods. We, therefore, envision that parameterization using just a single ionic conformation is in line with the selected model Hamiltonian. Further efforts to account for conformational flexibility in AAILs will greatly complicate development of the force field. There is no a priori warranty that multiple conformations included in the FF derivation algorithm will provide a higher accuracy for the properties of interest.

**Conclusions**

We have developed a new force field for the seven imidazolium-based amino acid ionic liquids featuring polar AA residues. The elaborated force field does not contain an explicit interaction term responsible for electronic polarization effects. Consequently, the model allows achieving a very high computational productivity and scalability (parallelization) on modern machines. Specific cation–anion interactions, including polarization, are on the average captured by the refined electrostatic potential. Unlike normally done, electrostatic potential has been derived using electronic structure description of a neutral ion pair, rather than of an isolated ion. The developed force field fosters numerical simulations of AAILs and promotes the field of room-temperature ionic liquids. Our results are also expected to inspire efforts of experimentalists to determine a wide range of properties of this emerging class of ionic liquids.

**Acknowledgments**

E.E.F. thanks Brazilian agencies FAPESP and CNPq for support. V.V.C. acknowledges research grant from CAPES (Coordenação de Aperfeiçoamento de Pessoal de Nível Superior, Brasil) under "Science Without Borders" program.

**Supporting Information**

The reported force field is covered by the GPL license. An interested reader can obtain ready-to-use files by e-mailing to fileti@gmail.com or vvchaban@gmail.com. Please, include your name and academic affiliation in the message. Alternatively, all the developed parameters are explicitly mentioned in the manuscript and can be, together with the CHARMM36 FF, used to construct topology files for the popular molecular dynamics simulation suites.

**AUTHOR INFORMATION**

E-mail addresses for correspondence: fileti@gmail.com (E.E.F.); vvchaban@gmail.com (V.V.C.)

TOC Graphic

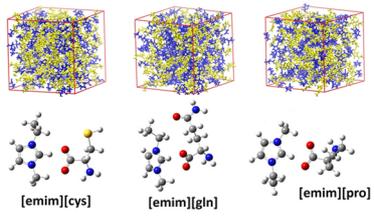